\documentclass[twocolumn,showpacs,preprintnumbers,amsmath,amssymb]{revtex4}

\usepackage{epsfig}
\usepackage{subfigure}
\usepackage{bm}

\usepackage[normalem]{ulem} 
\usepackage[dvips]{color} 
\renewcommand\sout{\bgroup \color{red} \ULdepth=-.5ex \ULset}

\newcommand{\be}{\begin{equation}}
\newcommand{\ee}{\end{equation}}
\newcommand{\bea}{\begin{eqnarray}}
\newcommand{\eea}{\end{eqnarray}}

\newcommand{\nn}{\nonumber\\}

\begin{document}

\title{Analysis of dilepton production in {Au+Au} collisions at $\sqrt{s{_{NN}}}=200$~GeV \\ within
the Parton-Hadron-String-Dynamics (PHSD) transport approach}

\author{O.~Linnyk}

\email{Olena.Linnyk@theo.physik.uni-giessen.de}

\affiliation{%
 Institut f\"ur Theoretische Physik, %
  Universit\"at Giessen, %
  35392 Giessen, %
  Germany %
}

\author{W.~Cassing}
\affiliation{%
 Institut f\"ur Theoretische Physik, %
  Universit\"at Giessen, %
  35392 Giessen, %
  Germany %
}

\author{J.~Manninen}%
\affiliation{%
 Institut f\"ur Theoretische Physik, %
 Johann Wolfgang Goethe University, %
 60438 Frankfurt am Main, %
 Germany; %
Frankfurt Institute for Advanced Studies, %
 60438 Frankfurt am Main, %
 Germany; %
}

\author{E.~L.~Bratkovskaya}%
\affiliation{%
 Institut f\"ur Theoretische Physik, %
 Johann Wolfgang Goethe University, %
 60438 Frankfurt am Main, %
 Germany; %
Frankfurt Institute for Advanced Studies, %
 60438 Frankfurt am Main, %
 Germany; %
}

\author{C. M.~Ko}%
\affiliation{%
Cyclotron Institute and Department of Physics and Astronomy, %
Texas A\&M University, %
College Station, TX 77843-3366, USA}

\date{\today}

\begin{abstract}

We address  dilepton production in Au+Au collisions at
$\sqrt{s{_{NN}}}=200$~GeV by employing the parton-hadron-string
dynamics (PHSD) off-shell transport approach. Within {the} PHSD  one
{goes beyond the quasiparticle approximation by solving} generalized
transport equations on the basis of the off-shell Kadanoff-Baym
equations for {the} Green's functions in {the} phase-space
representation. The approach consistently describes the full
evolution of a relativistic heavy-ion collision from the initial
hard scatterings and string formation through the dynamical
deconfinement phase transition to the quark-gluon plasma (QGP) as
well as hadronization and to the subsequent interactions in the
hadronic phase. {With partons described in the PHSD by the dynamical
quasiparticle model (DQPM) - matched to reproduce lattice QCD
results in thermodynamic equilibrium} - we calculate{, in
particular,} the dilepton radiation from  parton{ic} interactions
through the reactions $q\bar q\to\gamma^*$, $q\bar q\to\gamma^*+g$
and $qg\to\gamma^*q$ ($\bar q g\to\gamma^* \bar q$) in the early
stage of relativistic heavy-ion collisions. By comparing our results
to the data {from} the PHENIX  Collaboration, we study the relative
importance of different dilepton production mechanisms and point out
the regions in phase space where partonic channels are dominant.
Furthermore, explicit predictions are presented for dileptons within
the acceptance of the STAR detector system and compared to the
preliminary data.

\end{abstract}

\pacs{%
25.75.-q, 25.75.Cj, 25.75.Nq, 24.85.+p 
}

\keywords{%
Relativistic heavy-ion collisions\sep Meson production\sep
Quark-gluon plasma }

\maketitle

\section{Introduction}

The nature of confinement and the phase transition from a partonic
system of quarks, antiquarks and gluons{, the so-called} quark-gluon
plasma (QGP), to interacting hadrons, as occurring in relativistic
nucleus-nucleus collisions, is a central topic of modern high energy
physics. Already some decades ago dileptons {were} suggested as
useful probes of the {properties of the}
QGP~\cite{Shuryak:1977ut,Shuryak:1978ij,Feinberg:1970tg,Feinberg:1976ua,Bjorken:1975dk}.
Since dileptons are emitted over the entire history of the heavy-ion
collision, from the initial nucleon-nucleon collisions through the
hot and dense (partonic) phase and to the hadron decays after
freeze-out, microscopic transport models have to be applied for
disentangling the various sources that contribute to the final
dilepton spectra seen in experiments.

In the present work, we study dilepton production in {Au+Au}
collisions at $\sqrt{s{_{NN}}}=200$~GeV within the  PHSD off-shell
transport approach {by} including {the} collisional broadening of
vector mesons, microscopic secondary multi-meson channels{,}  and
the {radiation from the} strongly interacting QGP {through} the
interactions of dynamical quasiparticles having broad spectral
functions {in line with the degrees of freedom propagated} in the
transport approach.

The PHENIX Collaboration has presented dilepton data from $pp$ and
{Au+Au} collisions at {the} Relativistic-Heavy-Ion-Collider (RHIC)
{energy} of
$\sqrt{s{_{NN}}}$=200~GeV~\cite{PHENIXpp,PHENIX,PHENIXlast}, which
show a large enhancement relative to {the} scaled $pp$ collisions in
the invariant mass regime from 0.15 to 0.6 GeV/c$^2$. The question
that we aim at investigating in this work is whether {this} excess
can be attributed to the gluon Compton scattering or other partonic
interaction processes, which due to the two-particle finite state
contribute at low masses {as well}.

Moreover, if a realistic partial loss of the $D$- and $\bar D$-meson
correlations due to their rescattering is taken into account, one
has to conclude from the data of the PHENIX Collaboration that there
exists another domain of invariant mass{es}, in which the measured
dilepton yield in {Au+Au} collisions is underestimated by the scaled
yield from {the} $p+p$ {collisions}, i.e. at masses from 1 to
4~GeV/c$^2$~\cite{Linnyk:2010ar,Manninen:2010yf}. The microscopic
calculations within the parton-hadron transport approach here will
answer if this discrepancy can be accounted for by the partonic
sources of dileptons.

In Ref.~\cite{Linnyk:2011hz}
we have recently presented {results from the} PHSD for the dilepton
spectrum produced in {In+In collisions} at 158~A{$\cdot$}GeV {and}
compared {them} to the NA60 data~\cite{NA60,Arnaldi:2008er}. We
recall that by employing the hadron-string dynamics (HSD) transport
approach that does not include explicit partonic {contributions} to
low mass dilepton production in relativistic heavy-ion collisions,
it was shown in Ref.~\cite{Bratkovskaya:2008bf} that the NA60 data
for the invariant mass spectra {of} $\mu^+\mu^-$ pairs from In+In
collisions at 158 A$\cdot$GeV indicated an in-medium modification of
the $\rho$-meson according to the `melting' scenario \cite{NA60}.
The {more recent} PHSD calculations in Ref.~\cite{Linnyk:2011hz}{,}
which took into account the phase transition to and the radiation
from the partonic phase, confirmed the HSD results that the spectrum
at invariant masses in the vicinity of the $\rho$ peak was better
reproduced by the $\rho$ meson yield if a broadening of {its}
spectral function in the medium was assumed. On the other hand, the
spectrum at $M>1$~GeV was shown to be dominated by partonic sources.
Moreover, {the inclusion of the} partonic dilepton sources {made it
possible} to reproduce in {the} PHSD the effective temperature {or
the inverse slope parameter of the transverse momentum spectrum} of
dileptons in the intermediate mass {region}. Furthermore, for
dileptons of low masses ($M<0.6$~GeV), a sizable contribution {from}
partonic processes{,} particularly the quark annihilation with gluon
bremsstrahlung in the final state{,} was found, {and this provides}
another {possible} window for probing the properties of the sQGP.

In the present work, our previous findings will be reexamined by
comparison to the dilepton measurements at RHIC energies. Previous
HSD results for $e^+e^-$ pairs in {Au+Au} collisions in comparison
to the data from the PHENIX Collaboration~\cite{PHENIX,PHENIXlast}
were presented in Ref.~\cite{Bratkovskaya:2008bf}. Whereas the total
yield {is} quite well described in the {low-mass} region {from} the
pion Dalitz decay as well as around the $\omega$, $\phi$ and
$J/\Psi$ mass{es}, {the} HSD clearly underestimates the measured
spectra in the regime from 0.2 to 0.6 GeV by approximately a factor
of 5 for central Au+Au collisions. After including the in-medium
modification {of} vector mesons, we {obtain a total} spectrum which
is only slightly enhanced compared to the 'free' scenario {of using
the vector meson properties in the vacuum} (see Fig.~6 of
Ref.~\cite{Bratkovskaya:2008bf}). The low mass dilepton spectra from
 {Au+Au} collisions {obtained by}
the PHENIX Collaboration {at RHIC} are {thus} clearly underestimated
in the invariant mass {region of} 0.2 to 0.6 GeV in the 'collisional
broadening' scenario as well as in the 'dropping mass + collisional
broadening' model. We mention that {the} HSD results for the
hadronic production of low mass dileptons are very close to {those}
calculated {by} van Hees and Rapp as well as {by} Dusling and
Zahed~\cite{Dusling} (cf. the comparison in
Refs.~\cite{PHENIXlast,AToia}).

At higher masses from 1 to 4~GeV, the dominant hadronic sources of
 lepton pairs are {from} the  semi-leptonic decays of correlated D-mesons and the dilepton
decays of charmonia. In {estimating this contribution, the
experimental information was used} in Ref.~\cite{Manninen:2010yf}
{to determine} the yields of charmed hadrons. Additionally, the
{effect} of $D$ meson rescattering {in the hadronic matter} was
{included in} the HSD transport
approach~\cite{Linnyk:2008hp,Linnyk:2007zx} to estimate the
{probability} of surviving correlated semileptonic charm decays in
heavy-ion collisions. {The} resulting {dilepton} yield {between the
$\phi$ and $J/\Psi$ peaks - after including} the semi-leptonic
decays of correlated D-mesons in {the} HSD - {was found to}
underestimate the PHENIX data by approximately a factor of
two~\cite{Manninen:2010yf}.

Another open question to be answered by the microscopic transport
calculations is the determination of `windows' in phase space for
observing dileptons from the quark-gluon plasma that possibly
overshine the hadronic sources. It has been originally suggested
that a substantial thermal yield from the deconfined phase might be
seen in the invariant mass region between the $\phi$ and $J/\Psi$
peaks~\cite{Shuryak:1978ij}, while the spectrum at lower masses was
dominated by meson decays. On the other hand, the calculations in
Ref{s}.~\cite{Linnyk:2011hz,Gallmeister:1999dj,Gallmeister:2000ra}
pointed to a possible second region of phase space for the
observation of the thermal QGP source at masses $\approx
0.3-0.6$~GeV. In order to clarify whether {these} dileptons of
masses $0.3-0.6$~GeV can be observed {among} the background of
dileptons from hadronic decays, a study within a transport approach
that incorporates dilepton production from the (non-equilibrium)
partonic phase, hadronic decays and the microscopic secondary
hadronic interactions -- including the ``$4 \pi$" channels -- thus
appears appropriate.

The {PHSD}~\cite{CasBrat,BrCa11} transport approach, which
incorporates the relevant off-shell dynamics of vector mesons and
the explicit partonic phase in the early hot and dense reaction
region as well as the dynamics of hadronization, allows for a
microscopic study of various dilepton {production} channels {in
non-equilibrium matter}. The PHSD off-shell transport approach is
particularly suitable for {investigating the enhanced production of
lepton pairs in the invariant mass range $0.3 \leq M \leq 0.7$
GeV/$c^2$ that is seen in experiments}, since it incorporates
various scenarios for the modification of vector mesons in a hot and
dense medium. In the present work, we calculate the spectra of
dileptons produced in the course of {Au+Au} collisions at
$\sqrt{s{_{NN}}}$=200~GeV from the partonic and hadronic sources by
including the partonic channels and the multi-meson channels besides
the usual hadron decay channels. By consistently treating in the
same microscopic transport framework both partonic and hadronic
phases of the {colliding} system, we are aiming to determine the
relative importance of different dilepton production mechanisms and
to point out the regions in phase space where partonic channels are
dominant.

The paper is organized as follows. In Sec.~\ref{section.PHSD}, we
give a brief description of the PHSD approach {as well as} the
hadronic and partonic sources of dilepton production incorporated in
{the} PHSD. {We then compare in} Sec.~\ref{section.results} results
{from our} calculations to the available experimental data for
{Au+Au collisions} at top RHIC energy. Finally, our conclusions are
presented in Sec.~\ref{section.conclusions}.



\section{The PHSD transport approach}
\label{section.PHSD}

The {PHSD}~\cite{CasBrat,BrCa11} {is an off-shell transport model
that} consistently describes the full evolution of a relativistic
heavy-ion collision from the initial hard scatterings and string
formation through the dynamical deconfinement phase transition to
the quark-gluon plasma as well as hadronization and to the
subsequent interactions in the hadronic phase. In the hadronic
sector, {the} PHSD is equivalent to the HSD  transport approach
\cite{Cass99,Brat97,Ehehalt} that has been used for the description
of $pA$ and $AA$ collisions from SIS to RHIC energies and has lead
to a fair reproduction of measured hadron abundances, rapidity
distributions and transverse momentum spectra. In particular, {as in
the} HSD{, the PHSD} incorporates off-shell dynamics for vector
mesons~\cite{Cass_off1} and a set of vector-meson spectral
functions~\cite{Brat08} that covers possible scenarios for their
in-medium modification{s}.
%
%
The transition from the partonic to hadronic degrees of freedom is
described by covariant transition rates for the fusion of
quark-antiquark pairs to mesonic resonances or three quarks
(antiquarks) to baryonic states, i.e.{, the} dynamical hadronization
\cite{Cass08}. Note that due to the off-shell nature of {both}
partons and hadrons, the hadronization process obeys all
conservation laws (i.e.{, the} 4-momentum conservation {and the}
flavor current conservation) in each event, the detailed balance
relations, and the increase in {the} total entropy $S$. The
transport theoretical description of quarks and gluons in {the} PHSD
is based on a dynamical quasiparticle model (DQPM) for partons {that
is} matched to reproduce {the} lattice QCD (lQCD) results {for a
quark-gluon plasma} in thermodynamic equilibrium. {The} DQPM
provides {the} mean-fields for gluons/quarks and {their} effective
2-body interactions {in the} PHSD. For details about the DQPM model
and the off-shell transport we refer the reader to
Ref.~\cite{Cassing:2008nn}.
%
%
We stress that a non-vanishing width $\gamma$ in the partonic
spectral functions is the main difference {between} the DQPM {and}
conventional quasiparticle models~\cite{qp1}. Its influence {on the
collision dynamics} is essentially seen in {the} correlation
functions, e.g., in the stationary limit{,} the correlation
{involving} the off-diagonal elements of the energy-momentum tensor
$T^{kl}$ defines the shear viscosity $\eta$ of the
medium~\cite{Peshier:2005pp}. Here a sizeable width is mandatory to
obtain a small ratio of the shear viscosity to entropy density
$\eta/s$, which results in a roughly hydrodynamical evolution of the
partonic system in PHSD \cite{Cass08}. The finite width leads to
two-particle correlations, which are taken into account by means of
the {\em generalized} off-shell transport
equations~\cite{Cass_off1}{,} that go beyond the mean field or
Boltzmann approximation~\cite{Cassing:2008nn,Linnyk:2011ee}.

\subsection{Partonic sources of dileptons in PHSD}

In the scope of the one- and two-particle interactions, dilepton
radiation by the constituents of the strongly interacting QGP
proceeds via following elementary processes:
the basic Born $q+\bar q$ annihilation mechanism, gluon Compton
scattering ($q+g\to \gamma^*+q$ and $\bar q+g\to \gamma^*+\bar
q$){,} and quark {and} anti-quark annihilation with {the} gluon
Bremsstrahlung in the final state ($q+\bar q\to g+\gamma^*$). In the
on-shell approximation, one would use perturbative QCD cross
sections for the processes listed above. However, in the strongly
interacting QGP the gluon and quark propagators differ significantly
from the non-interacting propagators. Accordingly, we have
calculated in Refs.~\cite{Linnyk:2004mt,olena2010} the off-shell
cross sections for dilepton production in the partonic channels by
off-shell partons, using the phenomenological parametrizations (from
the DQPM) for the quark and gluon propagators and their interaction
strength.

In Refs.~\cite{Linnyk:2004mt,olena2010} it was shown that the finite
quark and gluon masses modify the magnitude as well as the $M-$ and
$p_T-$dependence of the cross sections of the {above mentioned}
processes compared to the perturbative results for massless partons
(cf. Figs.~3 and 4 of Ref.~\cite{olena2010}). The modifications are
{large} at lower $M^2$ and at the edges of the phase space. It was
shown that the most prominent effect of the quark masses on the
dimuon production cross sections in the Born mechanism ($q+\bar q\to
\gamma^*$) was a sharp threshold value for the invariant mass of the
dilepton pair $M_{min}=m_1+m_2$. On the other hand, the finite
masses of the quark and antiquark produce additional higher-twist
corrections to the cross section, which decrease with increasing
$M^2$, so that the off-shell cross sections approach the leading
twist -- on-shell -- result in the limit of high dilepton masses. In
Fig.~4 of Ref.~\cite{olena2010}, an analogous comparison for the
$2\to2$ process $q+\bar q\to \gamma^*+g$  was shown by plotting the
off-shell (i.e. with finite masses for the quarks and gluons) cross
section for the quark annihilation with gluon bremsstrahlung in the
{final} state {of} various values of the quark and gluon
off-shellness (masses) and the corresponding on-shell result. As
found in Ref.~\cite{olena2010}, the maximum pair mass shifts to a
lower value {as a result of producing} a massive gluon in the final
state. For the rest of the $M$ values, the effect of the quark and
gluon masses is about 50\%.
%
%
For $m_{q/g}\to0$, the cross section approaches the leading twist
pQCD result.


The question of the effect of a finite parton width -- which
parametrizes  their interaction rate and correlation {as well as}
multiple scattering -- on dilepton rates in heavy-ion collisions has
been addressed in Ref.~\cite{olena2010} by convoluting the off-shell
cross sections with phenomenological spectral functions $A(m_q)$ and
$A(m_g)$ for the quarks and gluons in the quark-gluon plasma{,
respectively,} and with parton distributions in a heavy-ion
collision. The finite width of the quasiparticles was found to have
a sizable effect on the dilepton production rates. In particular,
the threshold of the Drell-Yan contribution was ``washed out". Also,
the shape and magnitude of the $2\to2$ processes ($q+\bar q\to
g+\gamma^*$ and $q+g\to q+\gamma^*$) {were} modified. One further
observed that the contribution of the gluon Compton process $q+g\to
q+\gamma^*$ to the rates was small compared to that of $q+\bar q$
annihilations.

In Ref.~\cite{Linnyk:2011hz}, we have implemented the cross sections
obtained in Refs.~\cite{Linnyk:2004mt,olena2010} into the PHSD
transport approach in the following way: Whenever the
quark-antiquark, quark-gluon and antiquark-gluon collisions occur in
the course of the Monte-Carlo simulation of the partonic phase in
{the} PHSD, a dilepton pair can be produced according to the
off-shell cross sections~\cite{olena2010}, which depend, in addition
to the virtualities of the partons involved, on the energy density
in the local cell {where} the collision takes place. The local
energy density governs the widths of the quark and gluon spectral
functions as well as the strong coupling in line with the DQPM.

\subsection{Hadronic sources of dileptons in PHSD}

In the hadronic sector, {the} PHSD is equivalent to the HSD
transport approach \cite{Cass99,Brat97,Ehehalt}. The implementation
of the hadronic decays into dileptons ($\pi$-, $\eta$-, $\eta '$-,
$\omega$-, $\Delta$-, $a_1$-Dalitz, $\rho\to l^+l^-$, $\omega\to
l^+l^-$, {and} $\phi\to l^+l^-$) in HSD (and PHSD) is described in
detail in Refs.~\cite{Brat08,Bratkovskaya:2008bf}. For the treatment
of the leptonic decays of open charm mesons and charmonia, we refer
the reader to
Refs.~\cite{Manninen:2010yf,Linnyk:2008hp,Linnyk:2007zx} for a
detailed description.

The PHSD off-shell transport approach is particularly suitable for
investigating the different scenarios for the modification of vector
mesons in a hot and dense medium. Just as the HSD model, {the} PHSD
approach incorporates the {\em off-shell propagation} for vector
mesons {as described in} Ref.~\cite{Cass_off1}. In the off-shell
transport{,} the hadron spectral functions change dynamically during
the propagation through the medium and evolve towards the on-shell
spectral function{s} in the vacuum.   As demonstrated in
Ref.~\cite{Brat08}, the off-shell dynamics is important for
resonances with a rather long lifetime in {the} vacuum but strongly
decreasing lifetime in the nuclear medium (especially $\omega$ and
$\phi$ mesons) and also proves vital for the correct description of
dilepton decays  of $\rho$ mesons with masses close to the two pion
decay threshold. For a detailed description of the off-shell
dynamics and the implementation of vector-meson modifications in
medium{,} we refer the reader to
Refs.~\cite{Cass_off1,Brat08,Bratkovskaya:2008bf,Linnyk:2011ee,Linnyk:2011hz}.

In Ref.~\cite{Linnyk:2011hz}, the extension of the hadronic sources
{in the PHSD} for dilepton production to include secondary
multi-meson interactions by incorporating the channels $\pi \omega
\to l^+l^-$, $\pi a_1\to l^+l^-$, {and} $\rho \rho \to l^+l^-$ is
described in detail. These so-called `4$\pi$ channels' for dilepton
production are incorporated in {the} PHSD on a microscopic level
rather than assuming thermal dilepton production rates and
incorporating a parametrization for the inverse reaction $\mu^+ +
\mu^- \rightarrow 4 \pi's$
by employing {the} detailed balance as in
{Refs.}~\cite{RH:2008lp,Santini:2011zw}.
By studying the electromagnetic emissivity (in the dilepton channel)
of the hot hadron gas, it was shown in
Refs.~\cite{Song:1994zs,Gale:1993zj} that the dominating hadronic
reactions contributing to the dilepton yield at the invariant masses
above the $\phi$ peak are the two-body reactions {of}
$\pi+\rho$, $\pi+\omega$, $\rho+\rho$, {and} $\pi+a_1$. This
conclusion was supported by the subsequent study in a hadronic
relativistic transport model~\cite{GLi}. Therefore, we {have}
implemented the above listed two-meson dilepton production channels
in the PHSD approach. In addition, some higher vector mesons
($\rho^\prime$ {\it etc.}) are tacitly included by using
phenomenological form factors {that are} adjusted to {the
experimental} data (cf. Ref.~\cite{Linnyk:2011hz}). {Specifically,
we} determined the cross sections for the mesonic interactions with
dileptons in the final state using an effective Lagrangian approach,
following the works of Refs.~\cite{Song:1994zs,GLi}. In order to fix
the form factors in the cross sections for dilepton production by
the interactions of $\pi+\rho$, $\pi +\omega$, $\rho+\rho$ and $\pi
a_1$, we used the measurements in the detailed-balance related
channels: $e^+e^-\to \pi+\rho$, $e^+e^-\to \pi +\omega$, $e^+e^-\to
\rho+\rho$, and $e^+e^-\to \pi+a_1$. Note that we fitted the form
factors while taking into account the widths of the $\rho$ and $a_1$
mesons in the final state by convoluting the cross sections with the
(vacuum) spectral functions of these mesons in line with
Ref.~\cite{Song:1995wy} (using the parametrizations of the spectral
functions as implemented in {the} HSD and described
in~\cite{Bratkovskaya:2008iq}). In Fig.~5 of
Ref.~\cite{Linnyk:2011hz} we have presented the resulting cross
sections, which are implemented in {the} PHSD.


\section{Comparison to data}
\label{section.results}

Let us first note that the bulk properties of heavy-ion {collisions}
at the top RHIC energy, such as the number of charged particles as
well as their rapidity, $p_T$, $v_2$ and transverse energy
distributions, are rather well described by {the} PHSD; {and} we
refer to Ref.~\cite{BrCa11} for an extended and detailed comparison
to the data from the PHOBOS, STAR and PHENIX collaborations. Since
the lQCD equation of state employed in {the} PHSD has a crossover
transition, the PHSD calculations show a rather long QGP phase in
central {Au+Au} collisions at $\sqrt{s{_{NN}}}=200$~GeV (cf. Fig.~6
of Ref.~\cite{BrCa11}) with the partonic degrees of freedom
dominating for about 5-7~fm/c. We recall that dilepton production in
the elementary $pp$ channel is {also} well under control in {the}
PHSD as {was previously} demonstrated in
Ref.~\cite{Bratkovskaya:2008bf}.

\begin{figure}
    \includegraphics[width=0.48\textwidth]{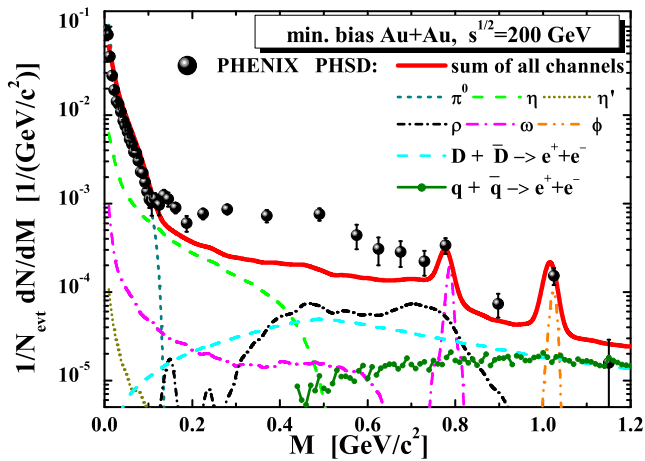}
    \caption{The PHSD results for the invariant mass spectra of
inclusive dilepton{s} in Au+Au collisions at $\sqrt{s{_{NN}}}$ = 200
GeV within the PHENIX acceptance cuts
{[Eq.}(\protect\ref{phenacc}){]} as given in the main text in
comparison to the data from the PHENIX
Collaboration~\protect{\cite{PHENIX,PHENIXlast}} for invariant
masses $M \! = \! 0 \! - \! 1.2$~GeV. The different lines indicate
the contributions from different channels as specified in the
figure. } \label{PHSD1}
\end{figure}

In Fig.~\ref{PHSD1} we show our results for {the invariant mass
spectra of} inclusive {dileptons in} Au+Au {collisions in the
invariant mass region} $M=0-1.2$~GeV for the acceptance cuts on
single electron transverse momenta $p_{eT}$, pseudorapidities
$\eta_e$, azimutal angle $\phi_e$, and dilepton pair rapidity $y${,}
\bea \label{phenacc} p_{eT}>0.2 \mbox{ GeV}, \nn |\eta_e|<0.35, \nn
-3\pi/16 < \phi_e < 5\pi/16 , \  11\pi/16 < \phi_e < 19\pi/16, \nn
|y|<0.35. \eea
In this region, the dilepton yield in the PHSD is dominated by
hadronic sources and essentially coincides with the earlier HSD
result~\cite{Bratkovskaya:2008bf}. There is a sizeable discrepancy
between the PHSD calculations and the data from the PHENIX
Collaboration in the region of masses from 0.2 to 0.6 GeV. The
discrepancy is not amended {by the inclusion of} the radiation from
the QGP as well as from correlated $D$-meson decays, since the
latter contributions are `over-shone' by the radiation from hadrons
integrated over the entire evolution of the collision.

\begin{figure}
    \includegraphics[width=0.48\textwidth]{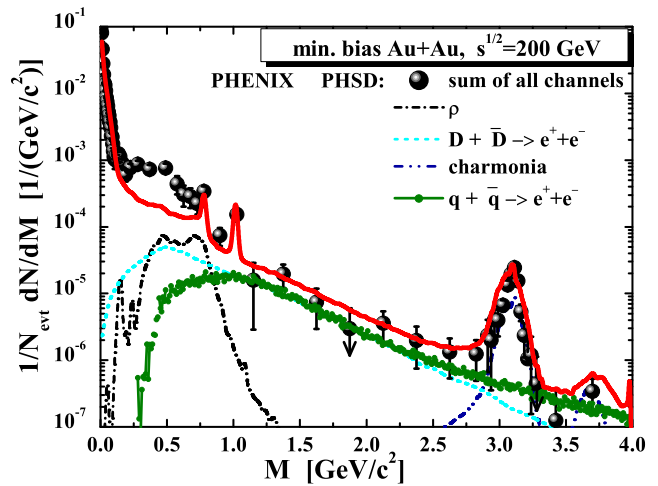}
     \caption{ {Same as Fig.~\ref{PHSD1}} for invariant masses
$M\!=\!0\!\,-\,\!4$~GeV.  } \label{PHSD2}
\end{figure}

In contrast, the partonic radiation as well as the yield  from
correlated $D$-meson decays are dominant in the mass region
$M=1-4$~GeV as seen in Fig.~\ref{PHSD2}, i.e.{,} in the mass region
between the $\phi$ and $J/\Psi$ peaks. The dileptons generated by
the quark-antiquark annihilation in the sQGP constitute about half
of the observed yield in this intermediate{-}mass range. For
$M>2.5$~GeV the partonic yield even dominates over the D-meson
contribution. Thus, {the inclusion of the} partonic radiation in
{the} PHSD fills up the gap between the hadronic model
results~\cite{Bratkovskaya:2008bf,Manninen:2010yf} and the data of
the PHENIX Collaboration for $M>1$~GeV. {Note that the collisional
broadening scenario for the modification of the $\rho$-meson was
used in the calculations presented in Figs.~\ref{PHSD1} and
\ref{PHSD2}.}

%
%

\subsection{Centrality dependence}

\begin{figure*}
\centering \subfigure[0-10\% centrality]{ \label{Bin1}
\resizebox{0.335\textwidth}{!}{%
 \includegraphics{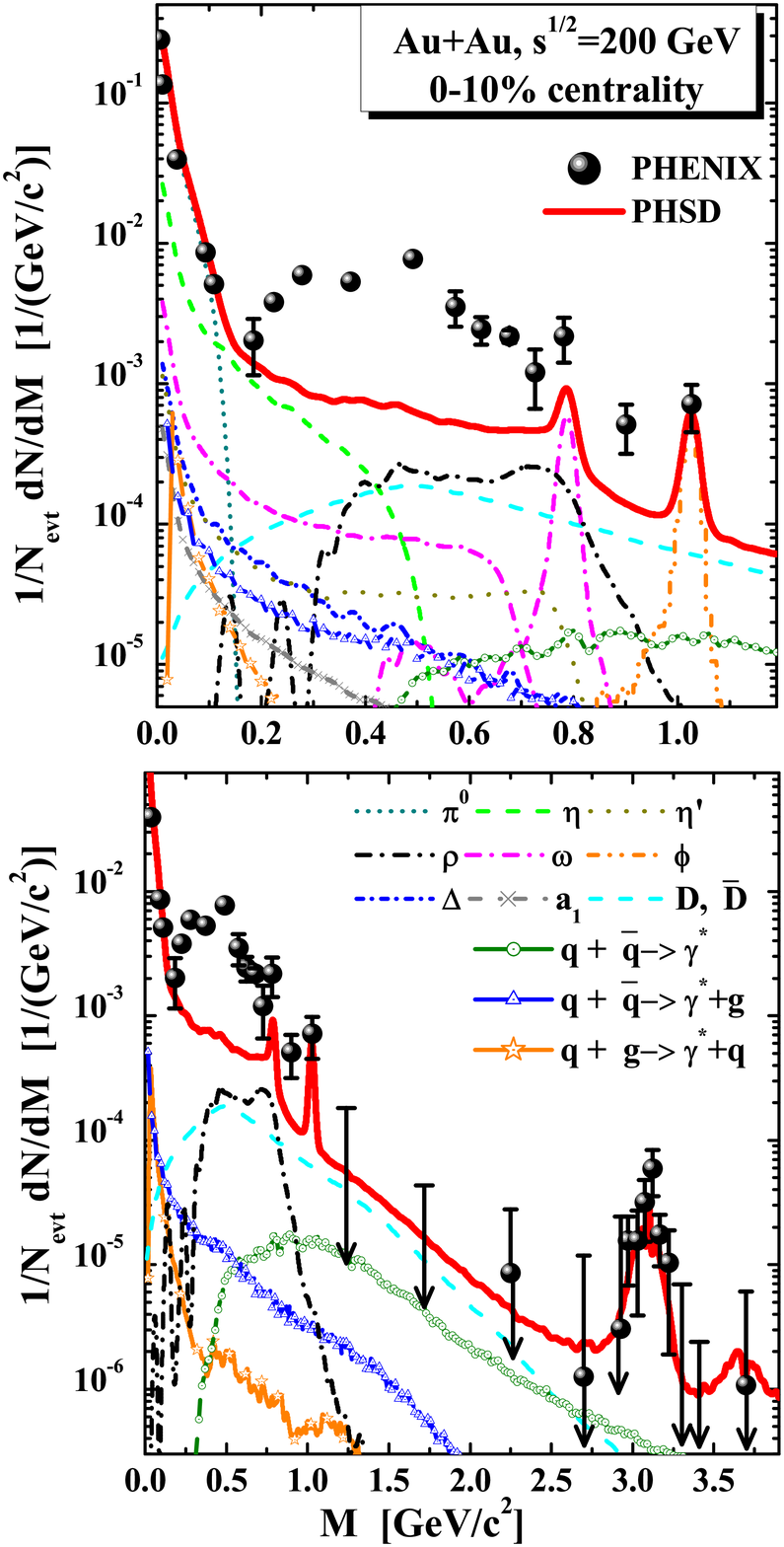}
} } \subfigure[10-20\% centrality]{ \label{Bin2}
\resizebox{0.305\textwidth}{!}{%
 \includegraphics{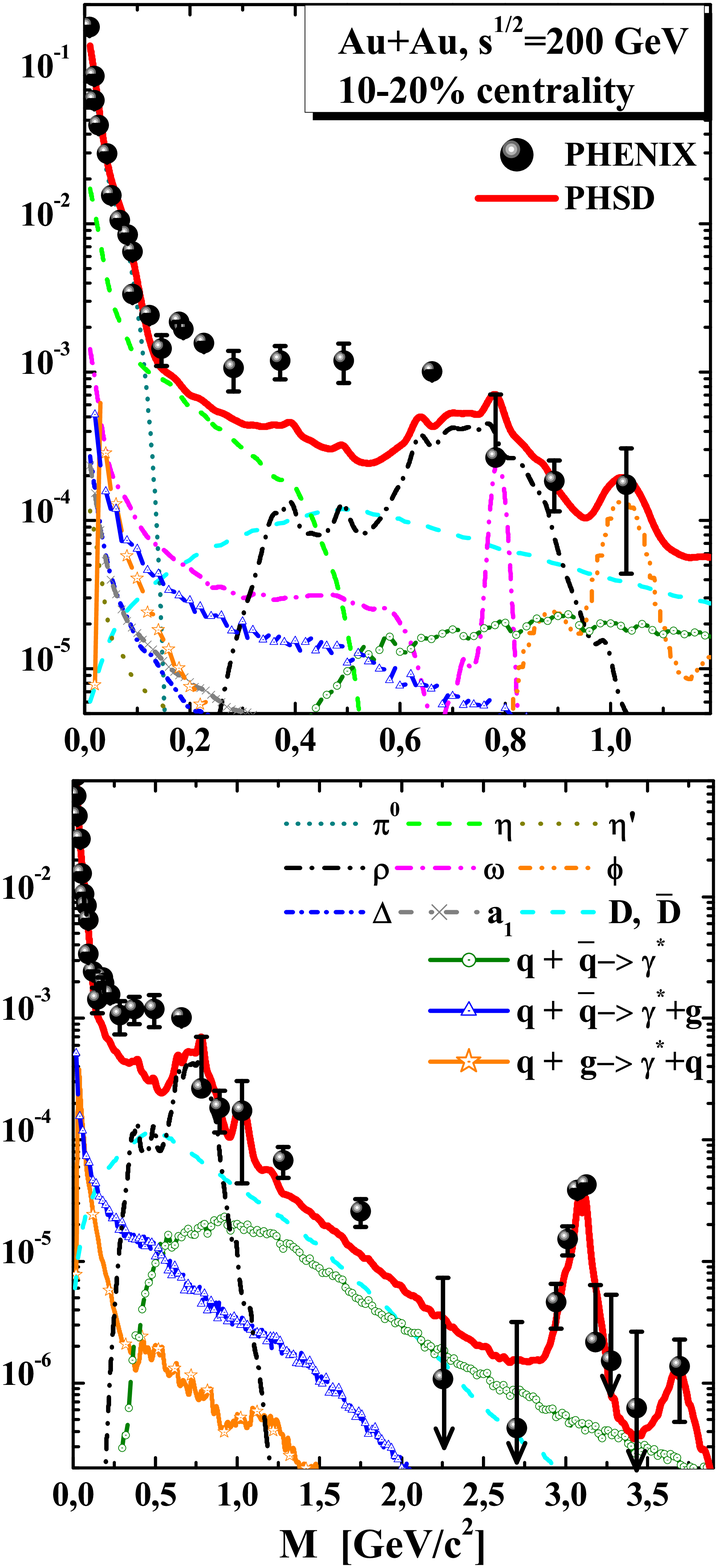}
} } \subfigure[20-40\% centrality]{ \label{Bin3}
\resizebox{0.305\textwidth}{!}{%
 \includegraphics{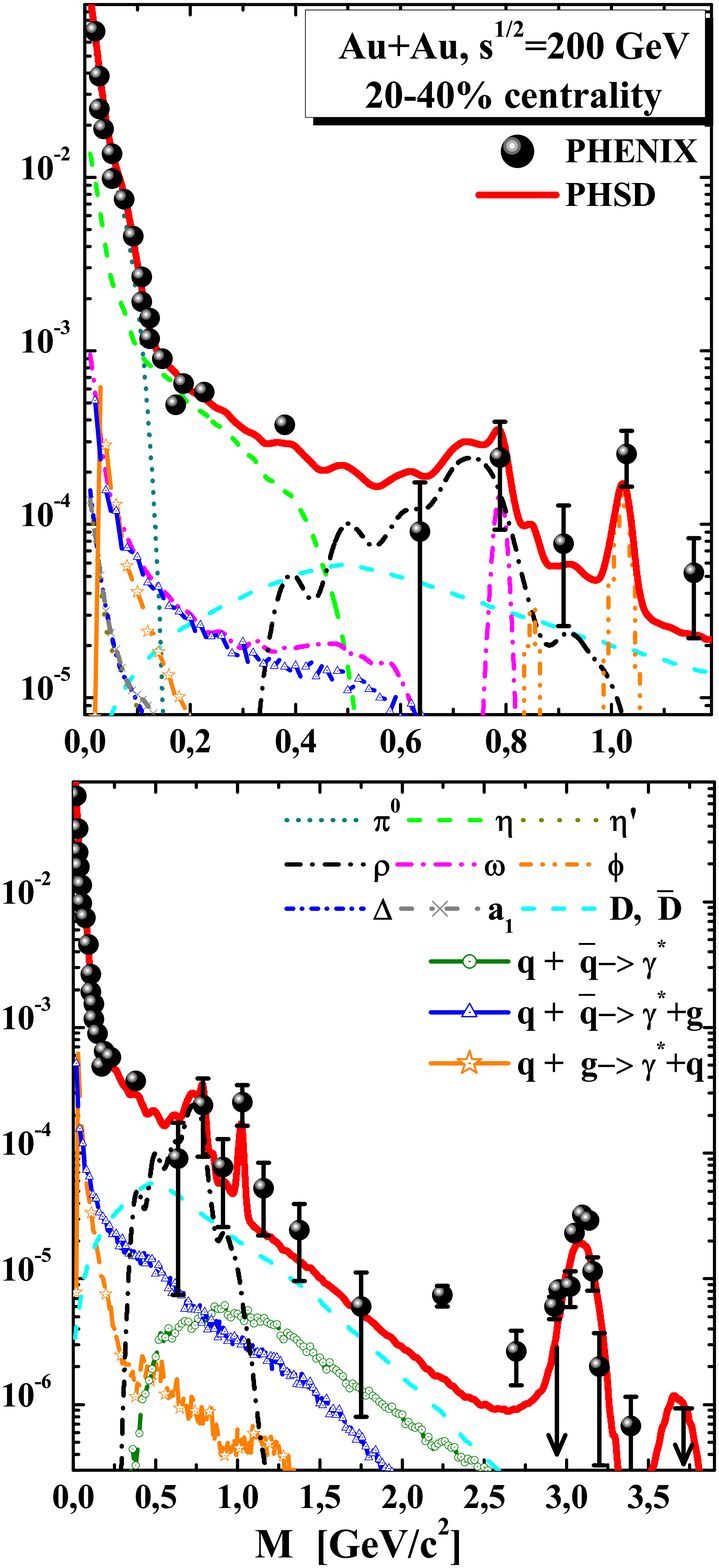}
} } \caption{ The PHSD results for the
 dilepton {invariant mass} spectra
 {in Au+Au} collisions of different
centralities at $\sqrt{s{_{NN}}}$ = 200 GeV in comparison to the
data from {the} PHENIX
{Collaboration}~\protect{\cite{PHENIX,PHENIXlast}}.}
\end{figure*}

Next we investigate the centrality dependence of dilepton production
in heavy-ion collisions at the top RHIC energy for the centrality
cuts specified in {Eq.}(\ref{phenacc}). While results from the PHSD
calculations are in a reasonable agreement with the PHENIX data at
20-40\% centrality{,} the data show an increasing excess {over the
PHSD results} for {the} 10-20\% centrality in the mass regime from
0.15-0.7 GeV{, and this excess becomes} even more dramatic for the
most central (0-10\% centrality) collisions (cf.
Figs.~\ref{Bin1}-\ref{Bin3}). For all these centrality cuts, the
contribution of partonic channels from {the} PHSD is subleading for
0.15 GeV $< M <$ 0.7 GeV and cannot be
 {responsible for} the excess
dileptons seen experimentally by the PHENIX Collaboration. In short,
the early expectation of a partonic signal in the low mass dilepton
spectrum is not verified by the microscopic PHSD calculations.

\subsection{Channel decomposition in central collisions}

\begin{figure}[h]
\includegraphics[width=0.48\textwidth]{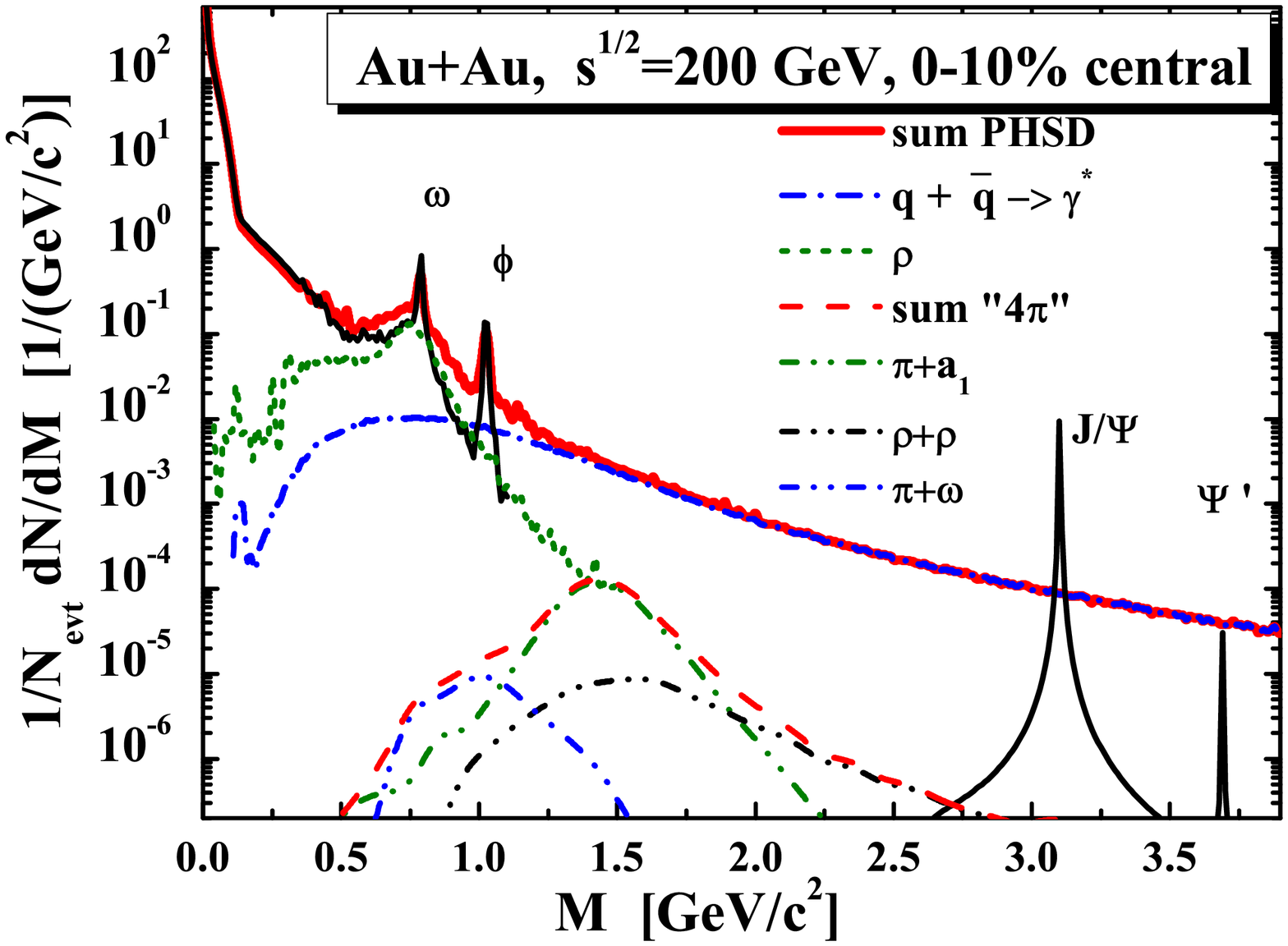}
\caption{Dilepton {invariant mass} spectra from central {Au+Au}
collisions at $\sqrt{s{_{NN}}}$ = 200 GeV integrated over rapidity
and transverse momenta as calculated in PHSD. The contributions from
multi-pion channels, quark-antiquark annihilation and hadron decays
into dileptons are shown separately.} \label{channels}
\end{figure}

In order to elucidate the relative importance of the different {\it
hadronic} sources of the excess dileptons in heavy-ion collisions at
the top RHIC energy, we show in Fig.~\ref{channels} the channel
decomposition of the main hadronic contributions to the dilepton
rates in central {Au+Au} collisions at $\sqrt{s{_{NN}}}=200$~GeV
without any cuts on dilepton momenta and rapidity and without
including the finite mass resolution of the PHENIX detector system.
Here the blue line with its sharp peaks from the decay of vector
mesons and a smooth background from the Dalitz decays of hadrons
becomes rather insignificant above the $\phi$-meson mass. Also the
$4 \pi$-channels ($\pi+a_1, \pi+\omega, \pi+\rho,$ {and} $
\rho+\rho$) are clearly subleading in the intermediate{-}mass
{region}. Here the partonic channels - dominated by $q + \bar{q}
\rightarrow e^+e^-$ - constitute about half of the dilepton yield
and {have} about the same contribution {as that} from correlated
$D$-meson decays (not shown in Fig.~\ref{channels}). Note that the
contribution from partonic channels is approximately exponential in
mass for 1 GeV $< M <$ 2.5 GeV and might be interpreted as being due
to 'thermal radiation' from the sQGP. However, the PHSD calculations
do not indicate that a thermal equilibrium has been achieved on the
partonic level.

\subsection{Transeverse momentum distributions}

\begin{figure}[h]
\includegraphics[width=0.47\textwidth]{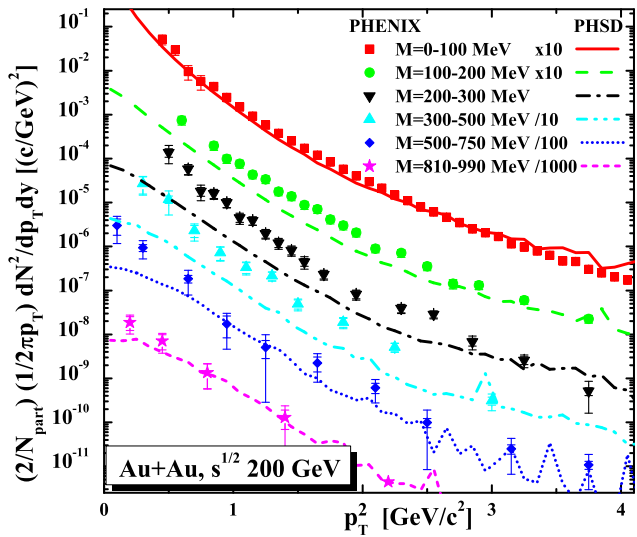}
\caption{The PHSD results for the transverse-momentum spectra of
dileptons from minimum bias {Au+Au}collisions at $\sqrt{s{_{NN}}}$ =
200~GeV in different mass bins compared to the data from {the}
PHENIX {Collaboration}~\protect{\cite{PHENIX,PHENIXlast}}. }
\label{PT_PHENIX}
\end{figure}

The PHENIX Collaboration has also accessed the information on the
transverse momentum dependence of dilepton production by measuring
the $p_T$-spectra of dileptons in different bins of invariant mass
$M$. In Fig.~\ref{PT_PHENIX} we show the measured transverse
momentum spectra of dileptons for minimum bias Au+Au collisions at
$\sqrt{s{_{NN}}}=200$~GeV (symbols) in comparison with the spectra
from {the} PHSD (lines) for six mass bins as indicated in the
figure. Whereas the {PHSD can well describe the} dilepton spectra in
the mass intervals [0,100 MeV] and [810 MeV, 990 MeV]{, it
underestimates the} low $p_T$ dileptons {in} the other mass bins{,
particularly} in the mass bins {[300 MeV, 500 MeV]}.  On the other
hand, high $p_T$ {dileptons} are reproduced quite well by the PHSD
calculations. We conclude that the missing dilepton yield for masses
from 0.15 to 0.6 GeV is essentially due {to} a severe
underestimation of the data at low $p_T$ by up to an order of
magnitude. We recall that at top SPS energies the low $p_T$ dilepton
yield could be attributed to $\pi - \pi$ annihilation channels, i.e.
to the soft hadronic reactions in the expansion phase of the system.
{These} channels are{, however,} insufficient to describe the very
low slope of the $p_T$ spectra at the top RHIC energy.

\subsection{Comparison with other models}

\begin{figure}[h]
\includegraphics[width=0.48\textwidth]{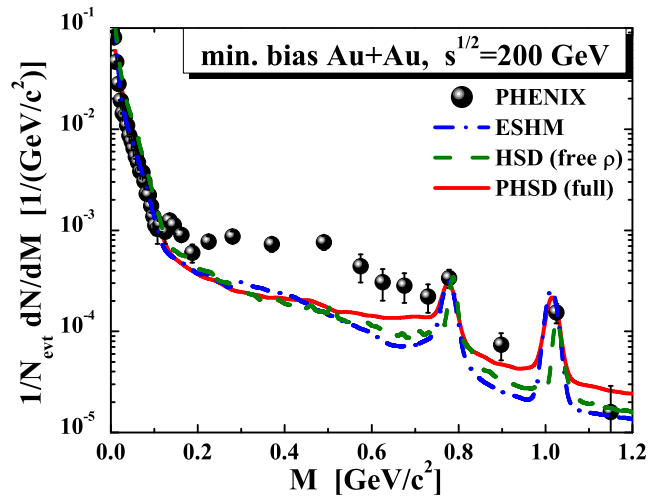}
\caption{{Invariant mass} spectra {of} inclusive {dileptons in
Au+Au} collisions at $\sqrt{s{_{NN}}}$ = 200 GeV in the low mass
region ($M \! = \! 0 \! - \! 1.2$~GeV) as calculated in three models
-- PHSD taking into account in-medium modification of rho {mesons
and} dilepton radiation from the partonic phase (solid line); HSD in
the free-rho scenario (dashed line); and the extended thermal
model~\protect{\cite{Manninen:2010yf}} -- compared to the data from
{the} PHENIX {Collaboration}~\protect{\cite{PHENIX,PHENIXlast}}. }
\label{Thermal}
\end{figure}

Since the authors have worked on the topic of dilepton production
within various approaches{,} it is instructive to discuss the
results especially in the low mass regime for the same centralities
and within the acceptance of the PHENIX detector system. In
Fig.~\ref{Thermal} we thus present the {invariant mass spectra of
inclusive} dilepton{s} {in Au+Au} collisions at $\sqrt{s{_{NN}}}$ =
200 GeV in the low mass region ($M \! = \! 0 \! - \! 1.2$~GeV) as
calculated within 1) PHSD (red solid line) taking into account {the}
in-medium modification of the $\rho$ as well as the dilepton
radiation from the partonic phase; 2) HSD in the free-$\rho$
scenario; and 3) the extended thermal model~\cite{Manninen:2010yf}
in comparison to the data from {the} PHENIX
{Collaboration}~\cite{PHENIX,PHENIXlast}.

The extended statistical hadronization model
(ESHM)~\cite{Manninen:2010yf} is an extension to the statistical
hadronization model (SHM) which has been applied
~\cite{SHM1,SHM2,SHM3,SHM4,SHM5,SHM6,SHM7,SHM8,SHM9,SHM10,SHM11,SHM12,SHM13,SHM14}
 to high-energy elementary and especially
heavy-ion collision experiments in order to calculate the yields of
different hadron species. {In the SHM, the} state of the ``thermal"
fireball is specified by its temperature $T$, volume $V$ and {the}
chemical potentials {$\mu_B$, $\mu_Q$, and $\mu_S$} for baryon,
electric and strangeness charges{, respectively}. {While $\mu_S$ and
$\mu_Q$ are zero} in central Au+Au collisions at
$\sqrt{s{_{NN}}}=200$~GeV, {$\mu_B$ is about  30 MeV.} The effect of
the strangeness under-saturation parameter {or fugacity} $\gamma_S$
on the di-electron invariant mass spectrum as a function of
centrality was studied in detail in {Ref.~\cite{Manninen:2010yf}}
and the effect was found to be moderate. We employ the value
$\gamma_S$=0.6 in this work for the min bias Au+Au collisions. The
overall normalization (fireball volume) at different centralities
was fitted to experimental data in {Ref.~\cite{Manninen:2010yf}} and
we use the same values throughout. For the temperature we use the
value $T = 170$~MeV.

{Since} the measured rapidity and transverse momentum spectra of
hadrons emitted in the high energy collision experiments do not
resemble thermal distributions{, the SHM has been extended in
Ref.~\cite{Manninen:2010yf}} {by boosting} (event by event) the
``fireball" along the beam axis so that the rapidity distributions
of pions become compatible with the BRAHMS
measurements~\cite{BRAHMS}. {Also, the} problem that the {SHM} tends
to over-populate the low $p_T$ part of the spectrum compared with
the experimental distributions was solved {in the ESHM} by assuming
that the created clusters' transverse momentum is normally
distributed with the width fitted together with the system volume
$V$ to the PHENIX data~\cite{PHENIXpiPT} in $p+p$ collisions and in
11 different centrality classes in the case of {Au+Au} collisions.
For further details we refer the reader to
Ref.~\cite{Manninen:2010yf}.

We find in Fig.~\ref{Thermal} that {the} HSD and the {ESHM} give
approximately the same results on the level of 30\% {for the
dilepton invariant mass spectra}. This might be surprising since
{the} HSD includes not only the direct and Dalitz decays of hadrons
but also meson-meson and meson-baryon channels for dileptons.
Indeed, the enhancement of the HSD result from 0.55 to 0.75~GeV can
be traced back to pion-pion annihilation which, however, gives only
a small contribution at the top RHIC energy. Our actual PHSD
calculations show some more dilepton yield in the $\rho$-mass regime
{as a result of} the broadened $\rho$ spectral function employed {in
the calculations}. In the 'free $\rho$' scenario, the results {from}
HSD and PHSD are identical within statistics since the partonic
channels give only a {small} contribution in this mass range. The
conclusion that the {dilepton} spectrum at masses below 1 GeV is
dominated by the hadronic sources is also supported by the studies
in other available models~\cite{Rapp:2010sj,Akamatsu:2011nr}.

\subsection{Predictions for STAR and comparison to preliminary data}

\begin{figure}
    \includegraphics[width=0.5\textwidth]{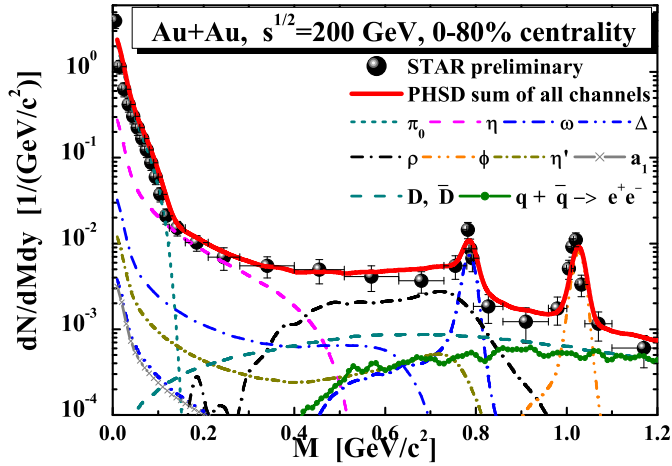}
    \caption{The PHSD results for the
{invariant mass} spectra of inclusive {dileptons in Au+Au}
collisions at $\sqrt{s{_{NN}}}$ = 200 GeV for $M \! = \! 0 \! - \!
1.2$~GeV and 0 - 80 \% centrality within the cuts of the STAR
experiment, see {[Eq.}(\protect\ref{staracc}){]} in the main text.
The preliminary data from the STAR Collaboration {are} adopted from
Ref. \protect \cite{Zhao:2011wa}.   } \label{STAR1}
\end{figure}

\begin{figure}
    \includegraphics[width=0.5\textwidth]{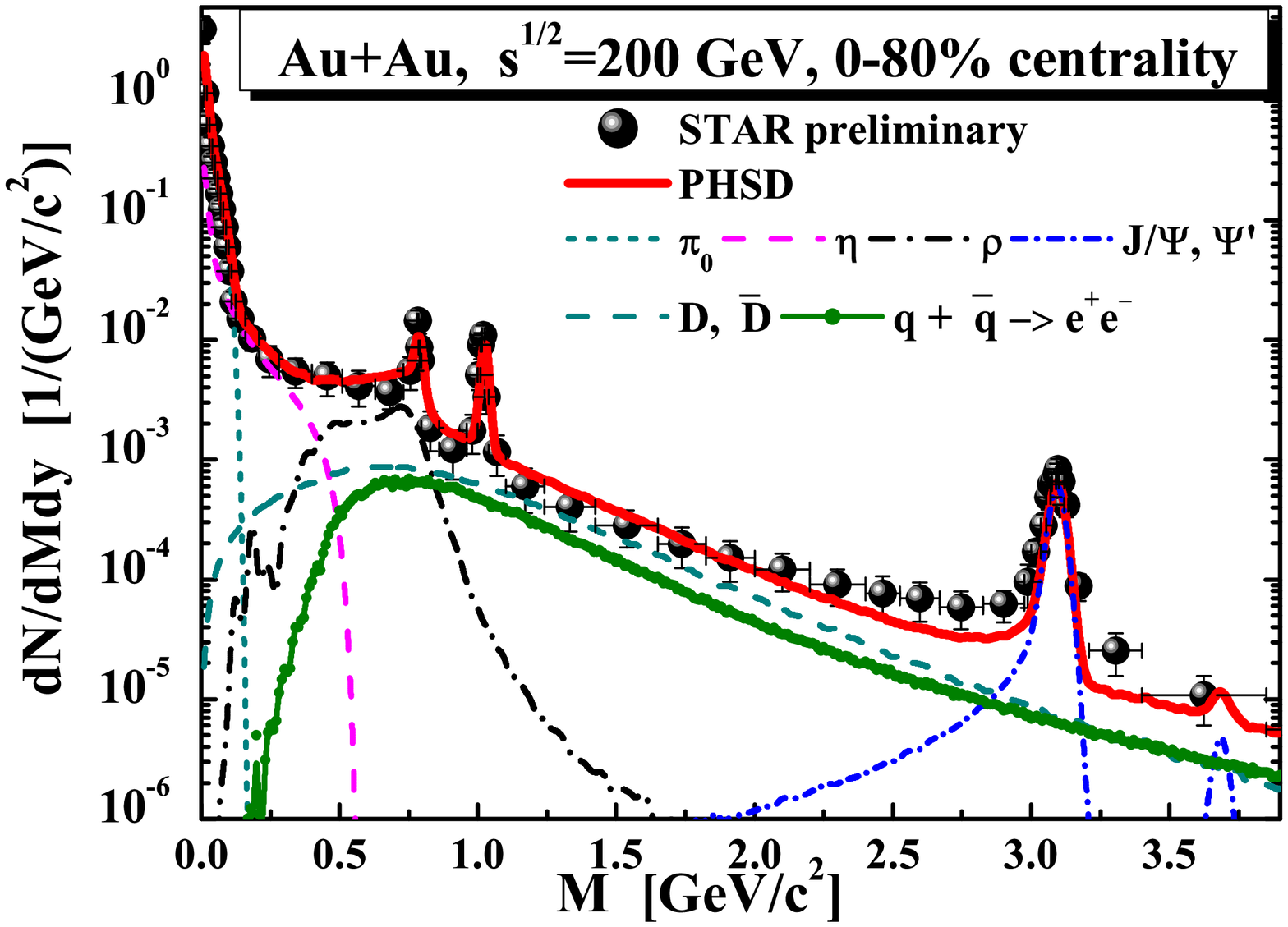}
     \caption{{Same as
Fig.~\ref{STAR1}} for $M\!=\!0\!\,-\,\!4$~GeV.} \label{STAR2}
\end{figure}

\begin{figure}
    \includegraphics[width=0.5\textwidth]{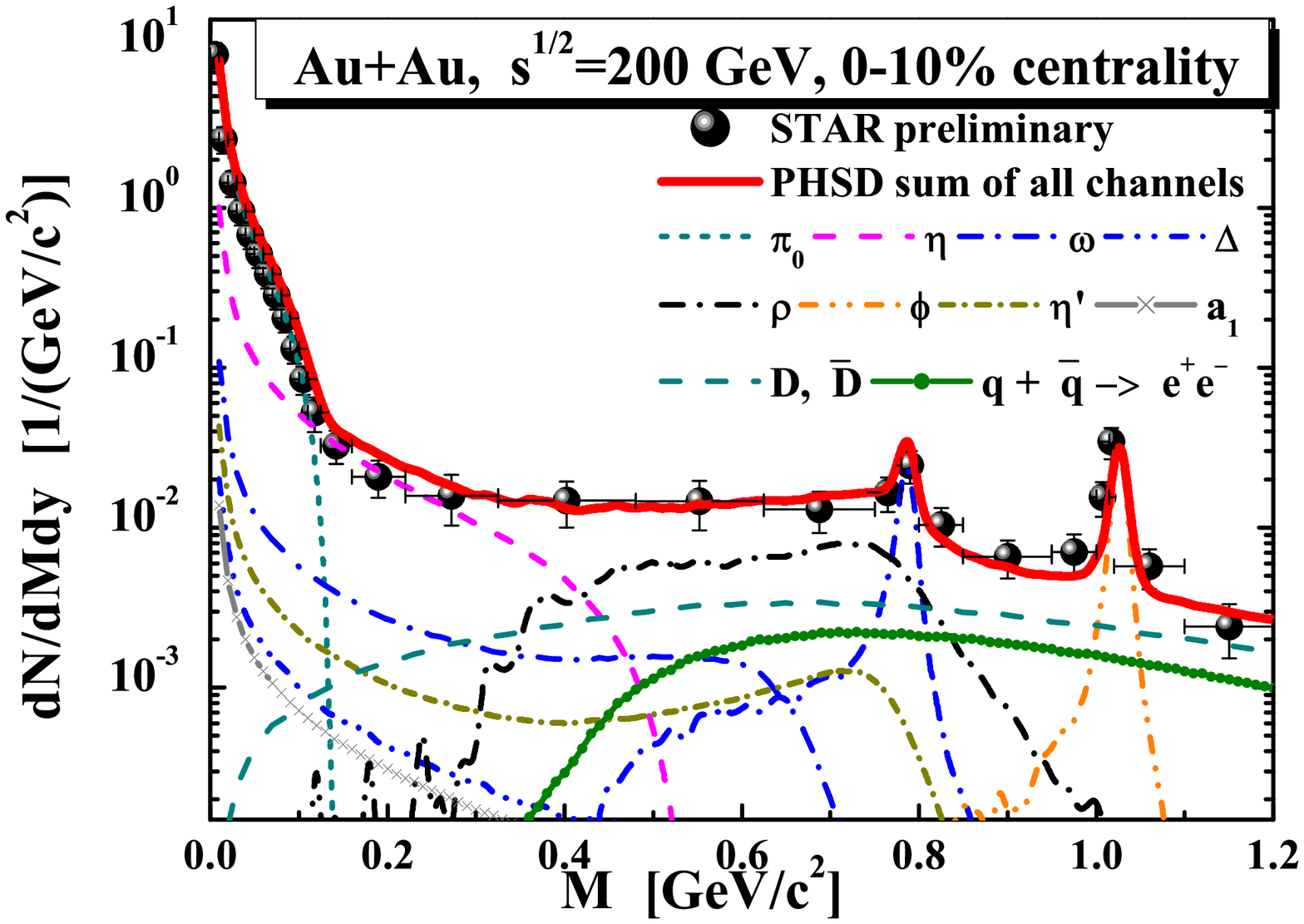}
    \caption{The PHSD results for the
{invariant mass} spectra of central {dileptons in Au+Au} collisions
at $\sqrt{s{_{NN}}}$ = 200 GeV for $M \! = \! 0 \! - \! 1.2$~GeV and
0 - 10 \% centrality within the cuts of the STAR experiment, see
Eq.[(\protect\ref{staracc})] in the main text. The preliminary data
from the STAR Collaboration {are} adopted from Ref.
\protect\cite{Zhao:2011wa}.  } \label{STAR3}
\end{figure}

\begin{figure}
    \includegraphics[width=0.5\textwidth]{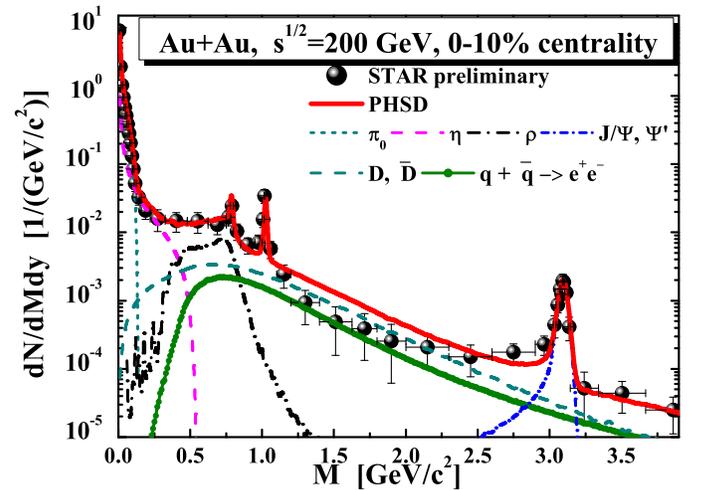}
     \caption{{Same as
Fig.~\ref{STAR3}} for $M\!=\!0\!\,-\,\!4$~GeV.} \label{STAR4}
\end{figure}

The PHSD calculations allow us to match with the different
experimental {conditions} and thus to provide a 'theoretical link'
between the different measurements. To this extent, we have provided
the differential data-tables for our theoretical predictions on the
web site~\cite{WebSTAR} so that any acceptance cuts and experimental
mass and transverse momentum resolutions can be applied.

The STAR Collaboration at RHIC has recently measured dileptons from
Au+Au collisions at $\sqrt{s{_{NN}}}$ = 200 GeV  with the acceptance
cuts on single electron transverse momenta $p_{eT}$, single electron
pseudorapidities $\eta_e$ and the dilepton pair rapidity $y$
\bea \label{staracc} 0.2<p_{eT}<5 \mbox{ GeV}, \nn |\eta_e|<1, \nn
|y|<1. \eea
Our predictions for the dilepton yield within these cuts are shown
in Figs.~\ref{STAR1} and \ref{STAR2} for 0-80\% centrality and in
Figs.~\ref{STAR3} and \ref{STAR4} for 0-10\% centrality. One can
observe generally a good agreement with the preliminary data from
the STAR Collaboration~\cite{Zhao:2011wa} for 0 - 80\% centrality in
the whole mass regime. Surprisingly, our calculations are also
roughly in line with the low mass dilepton spectrum from STAR in
case of central collisions whereas the PHSD results severely
underestimate the PHENIX data for the cuts given in {Eq.}(1) (cf.
Fig. 3). The observed yield from STAR can be accounted for by the
known hadronic sources, i.e. the decays of the $\pi_0$, $\eta$,
$\eta'$, $\omega$, $\rho$, $\phi$ and $a_1$ mesons, of the $\Delta$
particle and the semileptonic decays of the $D$ and $\bar D$ mesons,
where the collisional broadening of the $\rho$ meson is taken into
account! At first sight this observation might point towards an
inconsistency between the data sets from PHENIX and STAR but we have
to stress that the actual experimental acceptance cuts are more
sophisticated than those given in {Eqs.}(1) and (2). This problem
will have to be investigated closer by the experimental
collaborations. Furthermore, the upgrade of the PHENIX experiment
with a hadron blind detector~\cite{HBD} should provide decisive
information on the origin of the low mass dileptons produced in the
heavy-ion collisions at $\sqrt{s}=200$~GeV.

 We also observe a slight overestimation of the
dilepton yield from PHSD in 0-10\% central collisions at masses from
1.3 to 1.8~GeV where the dominant contributions to the spectrum  are
the radiation from the sQGP and the semileptonic decays of the $D$
and $\bar D$ mesons. We speculate that the suppression of dileptons
from the $D$ and $\bar D$ mesons might be underestimated in the PHSD
calculations in central collisions. The upgrade of the STAR
detector~\cite{Ruan:2009ug} will be promising in independently
measuring the correlated  $D$ and $\bar D$ meson contributions.


\section{Summary}

\label{section.conclusions}

In this study{,} we have addressed dilepton production in
 {Au+Au} collisions at $\sqrt{s{_{NN}}}=200$~GeV by
employing the Parton-Hadron-String Dynamics (PHSD) off-shell
transport approach. This work is  {a} continuation of our earlier
studies {for heavy ion collisions} at {the} SIS energies of 1-2 A
GeV \cite{Brat08} and {the} SPS energies from 40 to 158 A GeV
\cite{Linnyk:2011hz,Bratkovskaya:2008bf} essentially within the same
dynamical transport model. Within {the} PHSD one solves generalized
transport equations on the basis of the off-shell Kadanoff-Baym
equations for effective Green's functions in phase-space
representation (beyond the quasiparticle approximation) for quarks,
antiquarks and gluons as well as for the hadrons and their excited
states.  The PHSD approach consistently describes the full evolution
of a relativistic heavy-ion collision from the initial hard
scatterings and string formation through the dynamical deconfinement
phase transition to the quark-gluon plasma (QGP) as well as
hadronization and the subsequent interactions in the hadronic phase.
 {It was shown in} previous studies that the PHSD approach well describes the various hadron
abundancies, their longitudinal rapidity distributions as well as
transverse momentum distributions from lower SPS to top RHIC
energies~\cite{CasBrat,BrCa11}. Also the collective flow $v_2(p_t)$
is roughly in accordance with the experimental observations by the
PHOBOS, STAR and PHENIX collaborations at RHIC~\cite{BrCa11}. The
latter findings allow to explore the dynamics of subleading/rare
probes within the dynamical environment of partons and hadrons
during the complex time evolution of a relativistic heavy-ion
collision.

The present study has been devoted {particularly} to the calculation
of dilepton radiation from partonic interactions through the
reactions $q\bar q\to\gamma^*$, $q\bar q\to\gamma^*+g$ and
$qg\to\gamma^*q$ ($\bar q g\to\gamma^* \bar q$) in the early stage
of relativistic heavy-ion collisions at the top RHIC energy. We
recall that the differential cross sections for electromagnetic
radiation have been calculated with the same propagators as those
incorporated in the PHSD transport approach. By comparing our
calculated results to the data {from} the PHENIX Collaboration, we
have studied the relative importance of different dilepton
production mechanisms and addressed in particular the 'PHENIX
puzzle' of a large enhancement of dileptons in the mass range from
0.15 to 0.6 GeV as compared to the emission of hadronic states. Our
studies have demonstrated that the {observed} excess in the low mass
dilepton regime  cannot be attributed to partonic productions as
expected earlier. Thus the 'PHENIX puzzle' still has no explanation
from the theoretical approaches so far.
The PHENIX enhancement is essentially due to dileptons of low
transverse momentum in the mass range from 0.15 to 0.6 GeV and up to
date finds no explanation by hadronic nor partonic reaction channels
in PHSD that occur on top of the interactions in $pp$ collisions
during the evolution of relativistic heavy-ion collisions.

{Similar} to our findings at SPS energies \cite{Linnyk:2011hz}{,} we
find {that} the partonic dilepton production channels {are} visible
in the intermediate{-}mass {region} between the $\phi$ and $J/\Psi$
peaks. Their contribution is about the same as the correlated
background from $D$-meson decays. Surprisingly, this contribution
appears to be exponential in mass from 1 to 2.5 GeV {so} that an
interpretation of 'thermal radiation from the sQGP' might appear
appropriate. However, the PHSD dynamics shows that no kinetic
equilibrium is achieved on the partonic level in {heavy-ion
collisions} at top RHIC energies, and such an interpretation has to
be considered with care.

In view of the fact that within different (statistical and
dynamical) models we have not been able to find an explanation for
the low mass 'PHENIX puzzle'{,} the solution has to be relegated to
the experimental side. In fact, the STAR collaboration has taken
independent dilepton data for centralities different from the PHENIX
measurements and {also with} different detector acceptances. Our
PHSD calculations allow to match with the different experimental
{conditions} and thus to provide a 'theoretical link' between the
different measurements. To this extent, we have provided our
predictions~\cite{WebSTAR} for the conditions of the STAR
experiment, which happen to be in a rough agreement with the
preliminary data from the STAR Collaboration~\cite{Zhao:2011wa}.
This finding opens up new problems that will have to be addressed
from the experimental side. The upgrade of the PHENIX experiment
with a hadron blind detector~\cite{HBD} and the upgrade of the STAR
detector~\cite{Ruan:2009ug} for independently measuring the
correlated  $D$ and $\bar D$ meson contributions appear mandatory to
shed some further light on the present 'puzzles'.

\section*{Acknowledgements}

The authors are grateful for {the} fruitful discussions with
X.~Dong, T.~Hemmick, B.~Jacak, L.~Ruan, I.~Tserruya, A.~Toia and
N.~Xu. {They also} acknowledge {the} financial support through the
``HIC for FAIR" framework of the ``LOEWE" program and the Deutsche
Forschungsgemeinschaft (DFG). The work of C.M.K was supported by the
U.S. National Science Foundation under Grants No. PHY-0758115 and
No. PHY-1068572, the US Department of Energy under Contract No.
DE-FG02-10ER41682, and the Welch Foundation under Grant No. A-1358.


\end{document}